\documentclass[aps,prb,preprint,groupedaddress,showpacs,amsmath,amssymb]{revtex4}

\usepackage{graphicx}
\usepackage{dcolumn}
\usepackage{bm}

\begin{document}

\title{Specific heat jump at the superconducting transition temperature in Ba(Fe$_{1-x}$Co$_x$)$_2$As$_2$ and Ba(Fe$_{1-x}$Ni$_x$)$_2$As$_2$ single crystals}

\author{Sergey L. Bud'ko, Ni Ni, and Paul C. Canfield}
\affiliation{Ames Laboratory US DOE and Department of Physics and Astronomy, Iowa State University, Ames, IA 50011, USA}

\date{\today}

\begin{abstract}
We present detailed heat capacity measurements for Ba(Fe$_{1-x}$Co$_x$)$_2$As$_2$ and Ba(Fe$_{1-x}$Ni$_x$)$_2$As$_2$ single crystals in the vicinity of the superconducting transitions. The specific heat jump at the superconducting transition temperature ($T_c$), $\Delta C_p/T_c$, changes by a factor $\sim 10$ across these series. The $\Delta C_p/T_c$ vs. $T_c$ data of this work (together with the literature data for Ba(Fe$_{0.939}$Co$_{0.061}$)$_2$As$_2$, (Ba$_{0.55}$K$_{0.45})$Fe$_2$As$_2$, and (Ba$_{0.6}$K$_{0.4})$Fe$_2$As$_2$) scale well to a single log-log plot over two orders of magnitude in $\Delta C_p/T_c$ and over about an order of magnitude in $T_c$, giving $\Delta C_p/T_c \propto T_c^2$.
\end{abstract}


\maketitle

The discovery of superconductivity in F-doped LaFeAsO \cite{kam08a} and K-doped BaFe$_2$As$_2$ \cite{rot08a} compounds resulted in a large number of experimental and theoretical studies of the materials containing Fe-As layers as a structural unit. The understanding that transition metal (TM) substitution for Fe in the (AE)Fe$_2$As$_2$ series (AE = Ba, Sr, Ca) could be used to both stabilize superconductivity \cite{sef08a,lei08a,kum08a} and simplify growth while improving homogeneity, makes the (AE)(Fe$_{1-x}$TM$_x$)$_2$As$_2$ series (and in particular Ba(Fe$_{1-x}$TM$_x$)$_2$As$_2$ \cite{sef08a,ahi08a,nin08a,chu08a,lil09a,can09a}) a model system for studies of physical properties of Fe-As based materials.

Despite the large experimental and theoretical effort to understand the nature of superconductivity in Fe-As based materials, there are still a number of open issues that nave not been well resolved. From the experimental point of view, for number of properties, either the spread in the data is large, or systematic sets of data are absent. Temperature-dependent specific heat measured through the superconducting transition is one such property, even though it is considered to reflect of the superconductivity mechanism. Apart from the functional dependence of the specific heat ($C_p(T)$) below the superconducting transition temperature ($T_c$), the study of which often has complications caused by the need of careful substraction of the normal state contributions, the size of jump in $C_p$ at the superconducting transition is known to depend on the details of the superconducting state \cite{car90a,ska64a,mis03a,kog09a}, as judged and modeled by its deviation from the isotropic, weak coupling, BCS value of $\Delta C_p/\gamma T_c = 1.43$ ($\gamma$ being the normal state electronic specific heat).

The Ba(Fe$_{1-x}$Co$_x$)$_2$As$_2$ and Ba(Fe$_{1-x}$Ni$_x$)$_2$As$_2$ families of materials share very similar and complex $x - T$ phase diagrams\cite{nin08a,chu08a,can09a}:  initially on Co(Ni) -doping the critical temperature of the structural/antiferromagnetic transition decreases, with a separation in critical temperatures between these two transitions, \cite{pra09a,les09a} then, above some critical concentration, superconductivity is observed, apparently in the orthorhombic/antiferromagnetic phase. At higher Co(Ni) concentrations the structural/magnetic transitions are fully suppressed, whereas superconductivity appears to persist in the tetragonal phase up to $x \sim 0.14$ for TM = Co and $x \sim 0.08$ for TM = Ni. In this work we study the evolution of the specific heat jump at $T_c$ with TM = Co and Ni doping, for different concentration of TM. We examine the whole superconducting dome (both in orthorhombic/antiferromagnetic and tetragonal low temperature phases of the $x - T$ diagram) to gain insight into the details of the superconducting state in these materials.
\\

Single crystals of Ba(Fe$_{1-x}$Co$_x$)$_2$As$_2$ and Ba(Fe$_{1-x}$Ni$_x$)$_2$As$_2$ were grown out of self flux using conventional high-temperature solution growth techniques. \cite{can92a} Detailed description of the crystal growth procedure for this series can be found elsewhere. \cite{nin08a,can09a} The samples are plate-like with the plates being perpendicular to the crystallographic $c$-axis. The heat capacity data on the samples were measured using a hybrid adiabatic relaxation technique of the heat capacity option in a Quantum Design, PPMS instrument. Part of the $C_p(T)$ data for Ba(Fe$_{1-x}$Co$_x$)$_2$As$_2$ were presented, but not analyzed in detail, in Ref. \onlinecite{nin08a}.

\begin{figure}
\begin{center}
\includegraphics[angle=0,width=120mm]{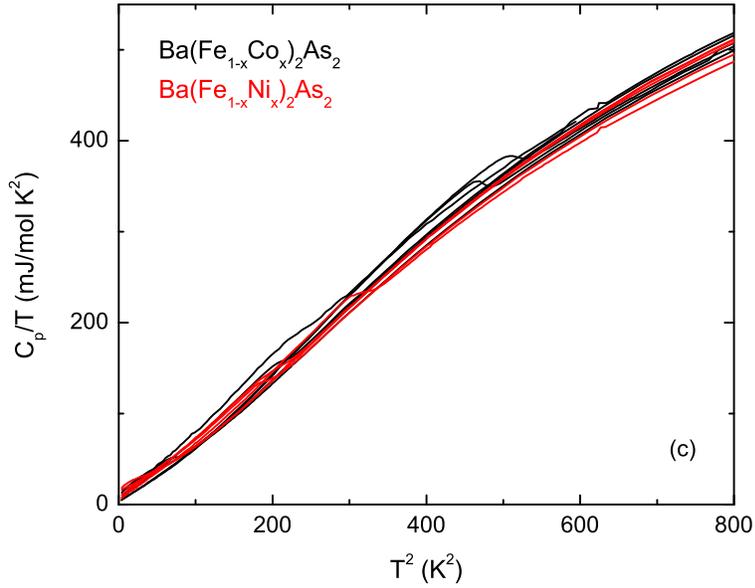}
\end{center}
\caption{(Color online) Temperature-dependent specific heat of (a) Ba(Fe$_{1-x}$Co$_x$)$_2$As$_2$, (b) Ba(Fe$_{1-x}$Ni$_x$)$_2$As$_2$ single crystals plotted as $C_p/T$ vs. $T$. Inset to panel (b): enlarged $C_p/T$ vs. $T$ plot near the superconducting transition for Ba(Fe$_{0.954}$Ni$_{0.046}$, lines show how $T_c$ and $\Delta C_p/T_c$ are estimated. Data in panels (a) and (b) are shifted by a multiple of 50 mJ/mol K$^2$ along the $y$ - axis for clarity. Panel (c): data for both series plotted as $C_p/T$ vs. $T^2$ without shifts.}\label{F1}
\end{figure}

It has to be noted, that both the superconducting transition temperatures and the upper critical fields in these materials are rather high \cite{nin08a}, thus making a reliable estimate of the normal state electronic specific heat, $\gamma$, difficult, especially bearing in mind that for approximately half of the samples in this study superconductivity coexists with an antiferromagnetic long range order. For this reason we are limited to the experimental determination of $\Delta C_p/T_c$, rather than the more traditional  quantity  $\Delta C_p/\gamma T_c$. Due to finite widths of the superconducting transitions, $\Delta C_p/T_c$ and $T_c$ values were determined from  plots of $C_p/T$ vs. $T$ using an "isoentropic" construction (Fig. \ref{F1}(b), inset). So defined values of $T_c$ may be slightly smaller than those reported for the samples from the same batches in Refs. [\onlinecite{nin08a,can09a}] in which different criterion was used.
\\

Temperature dependent heat capacity data for Ba(Fe$_{1-x}$Co$_x$)$_2$As$_2$ and Ba(Fe$_{1-x}$Ni$_x$)$_2$As$_2$, plotted as $C_p/T$ vs. $T^2$ are shown in Figs. \ref{F1}(a),(b). The jumps associated with the superconducting transitions are seen for all concentrations presented. Fig. \ref{F1}(c) presents all of the $C_p(T)$ data together showing that the spread of the of the data above the superconducting transitions is small, within 5-6\%, consistent with simple sample weighting errors. In addition, Fig. \ref{F1}(c) shows that both Co and Ni, added in these small amounts, are small perturbations to the BaFe$_2$As$_2$ system and do not significantly change the background $C_p$.

\begin{figure}
\begin{center}
\includegraphics[angle=0,width=120mm]{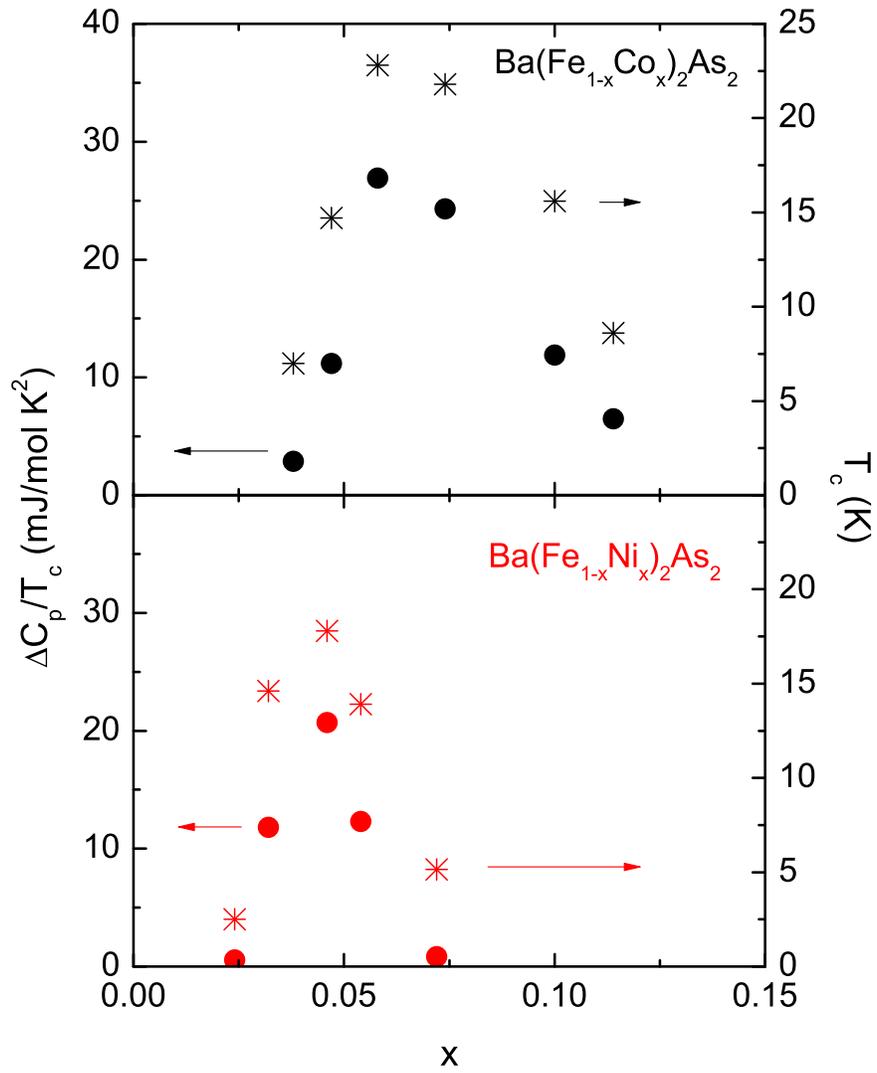}
\end{center}
\caption{(Color online) $\Delta C_p/T_c$ (circles, left axis) and $T_c$ (asterisks, right axis) as a function of concentration, $x$, Ba(Fe$_{1-x}$Co$_x$)$_2$As$_2$ (upper panel) and Ba(Fe$_{1-x}$Ni$_x$)$_2$As$_2$ (lower panel).}\label{F2}
\end{figure}

The values of the specific heat jumps at superconducting transition in the form of $\Delta C_p/T_c$ are plotted as a function of TM = Co, Ni concentration in Fig. \ref{F2}. The values of $T_c$ as a function of $x$ are displayed on the same plots as well. It is remarkable that (i) $\Delta C_p/T_c$ values change by as much as a factor of $\sim 10$ for the samples within the series; (ii) the shapes of the  $C_p/T_c$ vs. $x$ curves for both series appear to be related to the shapes of the respective $T_c$ vs. $x$ superconducting domes.

\begin{figure}
\begin{center}
\includegraphics[angle=0,width=120mm]{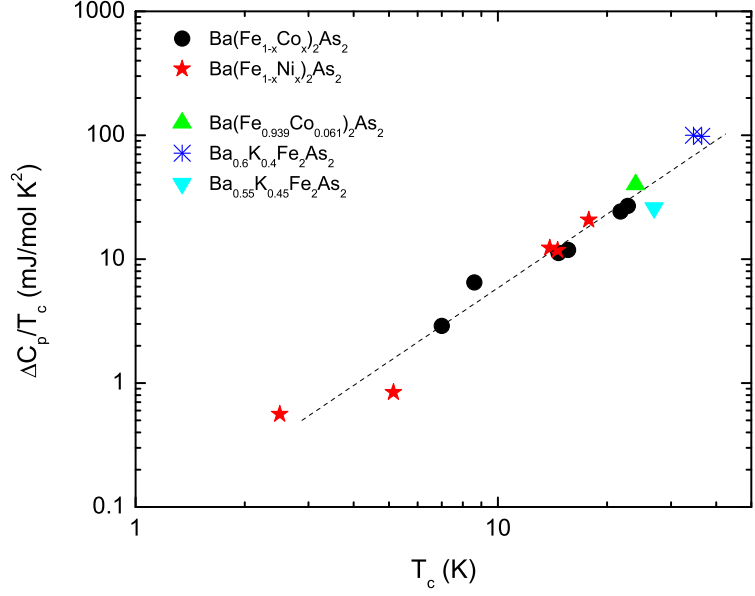}
\end{center}
\caption{(Color online) $\Delta C_p/T_c$ vs. $T_c$ for Ba(Fe$_{1-x}$Co$_x$)$_2$As$_2$ (circles) and Ba(Fe$_{1-x}$Ni$_x$)$_2$As$_2$ (stars) plotted together with literature data for Ba(Fe$_{0.939}$Co$_{0.061}$)$_2$As$_2$ [\onlinecite{chu08a}], Ba$_{0.55}$K$_{0.45}$Fe$_2$As$_2$ [\onlinecite{nin08b}], and Ba$_{0.6}$K$_{0.4}$Fe$_2$As$_2$ [\onlinecite{wel09a,mug09a}]. Dashed line has a slope $n = 2$ and is a guide for the eye.}\label{F3}
\end{figure}

These data can be plotted as $\Delta C_p/T_c$ vs. $T_c$ (Fig. \ref{F3}). It is curious, that all the data (both for "underdoped" and "overdoped" parts of the superconducting dome) collapse rather well onto a single curve. A data point \cite{chu08a} from another group's work on Ba(Fe$_{0.939}$Co$_{0.061}$)$_2$As$_2$ is consistent with our data, moreover, our previous result \cite{nin08b} on Ba$_{0.55}$K$_{0.45}$Fe$_2$As$_2$ as well as the very large $C_p$ jump at superconducting transition reported for Ba$_{0.6}$K$_{0.4}$Fe$_2$As$_2$ [\onlinecite{wel09a,mug09a}] follow the same trend (Fig. \ref{F3}). All the data in this figure can be fitted by a straight line with a slope $n \approx 2$.
\\

There are several possible ways to address such a remarkable behavior.

{\it Inhomogeneities}: it has been known that in a few cases (e.g. K-doping in BaFe$_2$As$_2$ samples \cite{nin08b,acz08a} the resulting samples had a distribution of dopant concentration, resulting in a broadening of the phase transitions. It appears to be less the case for TM doping: the wavelength dispersive x-ray spectroscopy does not show unambiguous, beyond the instrument error bars, distribution of TM dopant \cite{nin08a,can09a,nin09a}, for the more studied Ba(Fe$_{1-x}$Co$_x$)$_2$As$_2$ series, the low field susceptibility does not show variation in either transition width or magnetic flux expulsion, \cite{nin08a}  Meissner screening is homogeneous, \cite{gor09a}, structural and antiferromagnetic transitions for the intermediate concentrations are reasonably sharp \cite{nin08a,bud09a,pra09a}, and, for the phase diagrams and several other properties, results from the different groups appear to be very close. \cite{nin08a,chu08a} This being said, in an oversimplified model, inhomogeneities can be modeled by a uniform distribution of superconducting transition temperatures within some temperature  range. For an aggressive, $\pm 10$\% of $T_c$ spread, the width of the distribution (using $x = 0.074$ Co data as a starting point), the apparent $\Delta C_p/T_c$ jump will be approximately factor of 3 (but not $\sim 10$) smaller than the initial one, thus suggesting that inhomogeneity is not the sole reason for the observed behavior. In addition, the fact that the data in Fig. \ref{F3} are linear, with a slope $n \approx 2$, over two orders of magnitude in $\Delta C_p/T_c$ and over about an order of magnitude in $T_c$ on a log-log plot argues against an artifact caused by an uncontrolled spread in composition and may imply some more profound physical mechanism.

{\it Significant change of the density of states} within small, $\sim 10 \%$ range TM = Co, Ni doping, with $\gamma(x)$ or $DOS(x)$ having dome-like shape centered at $x$ values corresponding to the observed maximum in $T_c$: in this case $\Delta C_p/\gamma T_c$ could be close to constant or change insignificantly, whereas strongly $x$-dependent $\Delta C_p/T_c$ will be observed. Although, as mentioned above, reliable experimental data on normal state electronic specific heat as a function of $x$ are not available, band structure calculations on pure BaFe$_2$As$_2$ \cite{she08a,kre08a,liu08a} do not suggest such  a significant and {\it non-monotonic} change of the density of states for small TM = Co, Ni concentrations (a sharp local maximum in $\gamma$ would be required at optimal doping).

More physical reasons to observe significant change (decrease in comparison with isotropic, weak coupling, BCS case) in the heat capacity jump at $T_c$ could be associated with the effect of paramagnetic impurities, \cite{ska64a} multi-band  superconductivity, \cite{mis03a,kog09a} and/or effects of a normal state pseudogap. \cite{lor01a} Although each of these possibilities is plausible and exciting, it seems hard to construct a simple picture that will accommodate the observed dome-like, almost symmetric, $\Delta C_p/T_c$ vs. $x$ behavior within a single one of this models (unless a more complex case, in which left and right, "underdoped" and "overdoped", parts of the $\Delta C_p/T_c$ vs. $x$ dome, are explained separately by different mechanism, is considered).
\\

To summarize, approximate scaling of  $\Delta C_p/T_c$ with $T_c$ was observed for Ba(Fe$_{1-x}$Co$_x$)$_2$As$_2$ and Ba(Fe$_{1-x}$Ni$_x$)$_2$As$_2$ single crystals.  The reason for such scaling is not clear in this moment: if extrinsic (chemical inhomogeneities) it cannot be ignored in interpretation of other, detailed, experiments on these materials, if intrinsic, more work is required to elucidate the reason for this apparent scaling.

\begin{acknowledgments}

Work at the Ames Laboratory was supported by the US Department of Energy - Basic Energy Sciences under Contract No. DE-AC02-07CH11358. We thank Vladimir Kogan and J\"org Schmalian  for useful discussions and Jiaqiang Yan for help in synthesis.

\end{acknowledgments}

\clearpage

\begin{figure}
\begin{center}
\includegraphics[angle=0,width=120mm]{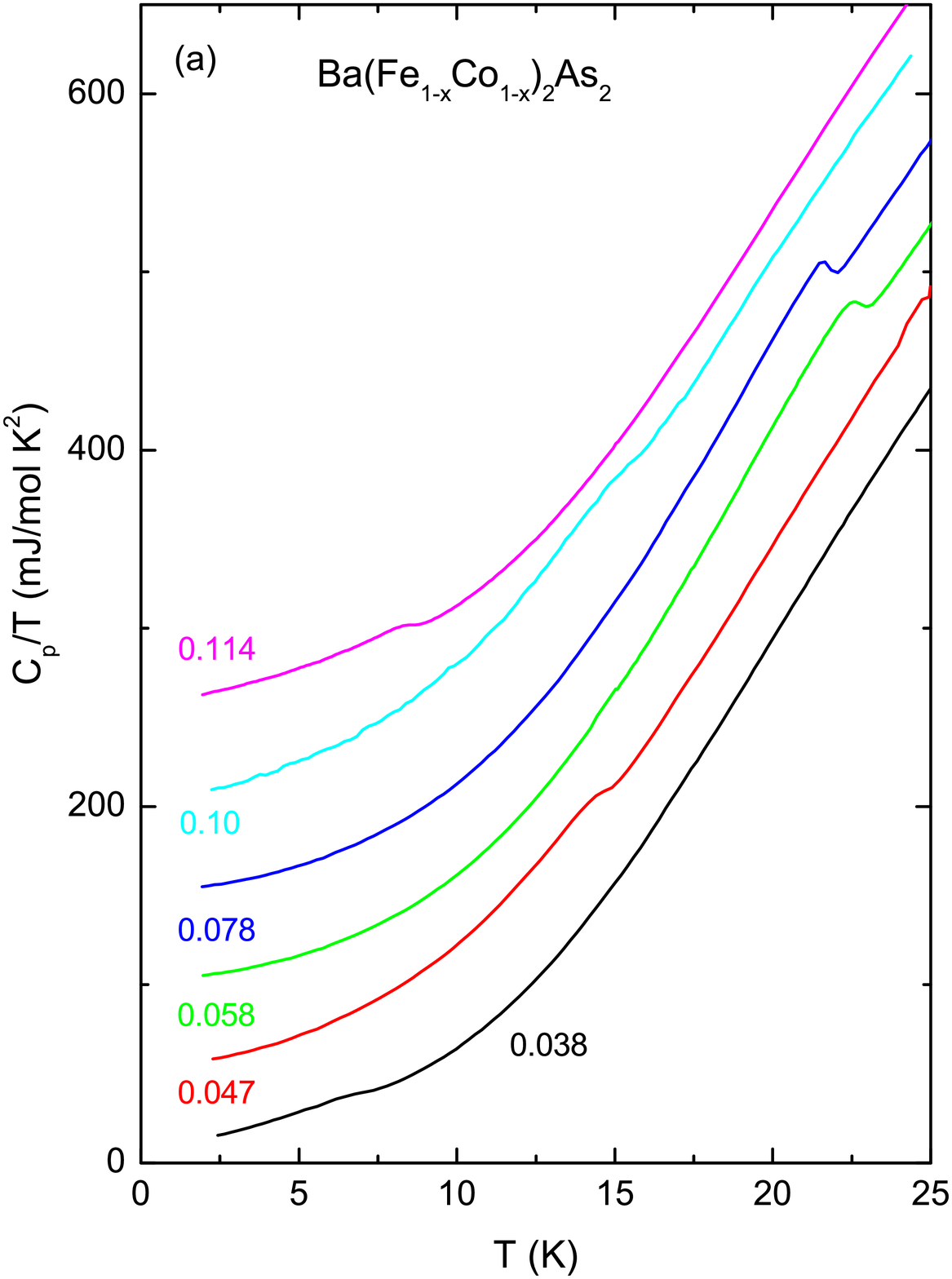}
\end{center}
\end{figure}

\clearpage

\begin{figure}
\begin{center}
\includegraphics[angle=0,width=120mm]{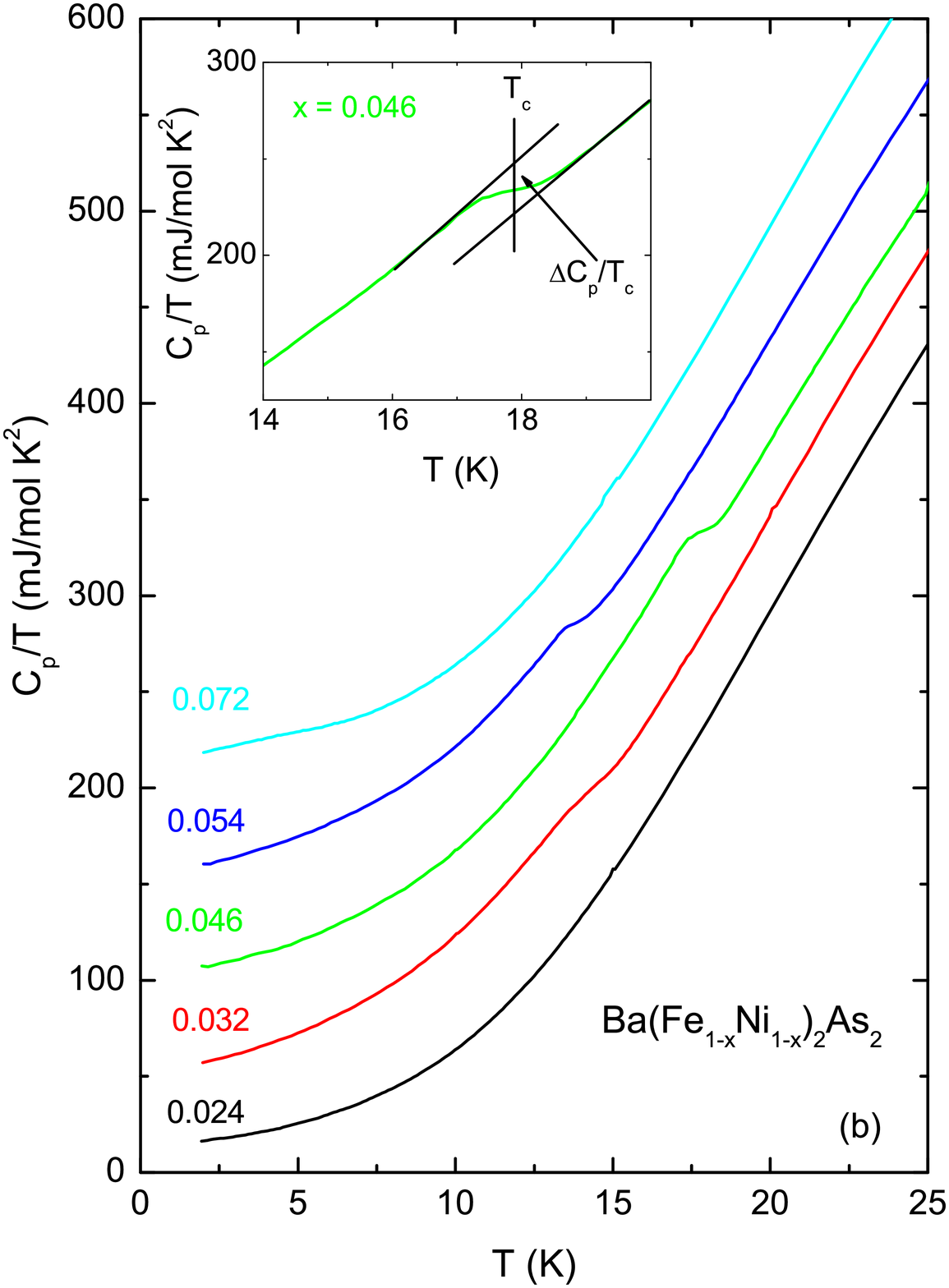}
\end{center}
\end{figure}

\clearpage

\clearpage

\clearpage

\end{document}